\documentstyle[11pt,newpasp,twoside,epsf]{article}
\input psfig.sty
\def\slHI{H{\,\small\sl I}}
\def\HI{H{\,\small I}}

\def\kms{km s$^{-1}$}

\def\edcomment#1{\iffalse\marginpar{\raggedright\sl#1\/}\else\relax\fi}
\reversemarginpar

\begin{document}
\title{HI absorption from the central kpc of
radio galaxies: effect of orientation or interstellar medium?}
\author{R. Morganti, T.A. Oosterloo}
\affil{Netherlands Foundation for Research in Astronomy, Postbus 2, NL-7990 AA
Dwingeloo, The Netherlands}
\author{C.N. Tadhunter, K.A. Wills}
\affil{University of Sheffield, Dept. of Physics, Sheffield, S3 7RH, UK}
\author{G. van Moorsel}
\affil{National Radio Astronomy Observatory, Socorro, NM 87801, USA}
\author{A. Capetti}
\affil{Osservatorio Astronomico di Torino, Strada Osservatorio 20, 10025 Pino
  Torinese, Italy}
\author{R. Fanti, P. Parma, H. de Ruiter}
\affil{Istituto di Radioastronomia, CNR, via Gobetti 101, 40129 Bologna, Italy}

\begin{abstract}
  We present a summary of recent studies of HI absorption in radio galaxies.
  The results show how the absorption can be due to a variety of phenomena and
  how they can help us in understanding more of what is happening in and around 
  AGNs.
\end{abstract}

\section{Introduction}

The study of neutral hydrogen is one of the tools we can use to investigate the
cool diffuse ISM in the central regions of AGNs and Starbursts.  This gas
component is believed to be connected to the fuelling and the obscuration
of the ``central engine'', both important aspects of the study of the AGNs. 
The infall of gas (possibly fuelling the AGN) can also trigger a central
starburst and give, perhaps, some clues of the AGN/starburst connection. 
Observations of HI absorption against radio loud AGNs have been carried out
for many years as soon as it was recognised that the radio activity might be
associated with the presence of cold gas (see e.g.  Gunn 1979).  
More recent observations - more sensitive and of higher resolution - have
shown how atomic gas can actually probe a variety of phenomena in the nuclear
regions around radio loud AGN.  To illustrate this, I will briefly summarise
some of the results we have recently obtained in the study of HI absorption.


\section{Thin disks in FRI radio galaxies?}

Obscuration is one of the main ingredients in the unification schemes of AGNs.
Thick tori are believed to be present in powerful radio galaxies and a
combination of beaming and obscuration would explain the lack of broad optical
emission lines in some of these galaxies.  From recent studies (both
observational e.g.\ Cygnus A, Conway 1998, and theoretical, Maloney et al.
1996), it is now clear that under certain conditions, the torus does not have
to be solely molecular, as original predicted, but atomic gas can be present
also in the very central regions of AGN, e.g. in the form of an obscuring
disk/torus-like structure.  For low power radio galaxies (i.e.  Fanaroff-Riley
type I), however, the situation is not so clear yet.  Using HST images,
Chiaberge et al.  (1999) found that unresolved optical cores are commonly
present in these radio galaxies.  A strong correlation is found between the
fluxes of the optical and the radio core, arguing for a common non-thermal
origin (synchrotron emission from the relativistic jet).  All this suggests
that the {\sl standard pc-scale geometrically thick torus is not present in
  these low-luminosity radio galaxies}.  Thin disks-like structures have been
observed in HI absorption in the case of NGC~4261 (van Langevelde et al.
2000) and Hydra~A (Taylor 1996). Thus, the study of HI absorption has the
potential to help us to answer the main question whether the FRI/FRII
dichotomy reflects fundamental differences in the innermost structure of the
central engine.



\begin{figure}[!h]
\centerline{\psfig{figure=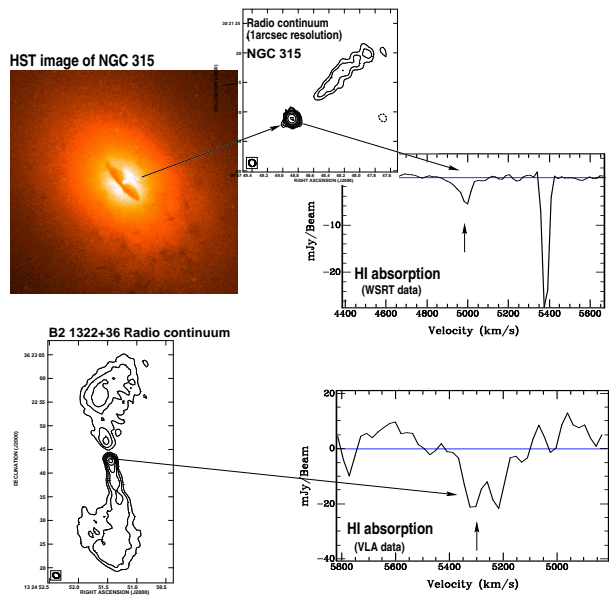,width=8cm,angle=-90}}
\caption{HI absorption detected with the WSRT and the VLA against 
the core of two radio  galaxies: NGC 315 and B2 1322+36}
\end{figure}

In a study of HI absorption of a complete sample of FRI radio galaxies
(Morganti et al.\ 2001), we have found a low rate of detections. Only one of
the 10 FRI galaxies observed were detected in HI absorption.  To first order,
this result is consistent with the idea that the cores of these radio galaxies
are relatively unobscured.
To investigate this idea in more detail, we have studied HI absorption (with
the VLA and the WSRT) in an other sample of radio galaxies for which
information from HST images about the presence of optical cores and nuclear
dusty disks/lanes are available (Capetti et al.\ 2000, de Ruiter et al. these
proceedings).  Thus, the HI observations {\sl aim to correlate the presence
(or absence) of HI absorption with the optical characteristics}.  We find HI
absorption in 4 of the 9 B2 galaxies observed.  In particular, absorption was
detected in the two galaxies in the sample that have dust disks/lanes and {\sl
no} optical cores (B2 1322+36 and B2 1350+31 (3C293)).  In these cases, the
column density of the absorption is quite high ($ > 10^{21}$ cm$^{-2}$ for
T$_{\rm spin} = 100$ K) and the derived optical extinction $A_B$ (between 1
and 2 magnitudes) is such that it can, indeed, produce the obscuration of the
optical cores. 
In the other two cases (B2 0055+30 (NGC 315) and B2 1346+26), HI absorption
has been detected despite the presence of optical cores.  The column density
derived from the detected absorption is, however, much lower ($\sim 10^{20}$
cm$^{-2}$ for T$_{\rm spin} = 100 $ K) and the derived extinction is of the
order of only a fraction of a magnitude.  The detected HI absorption could be
part (the innermost?) of the dusty disks seen with HST (see the one observed
in NGC 315 shown in Fig.\ 1) but the resolution of the VLA observations do not give
any spatial information (i.e.\ the HI absorption lines are all detected
against the unresolved core).  Only VLBI observations will be able to tell us
about the real distribution and kinematics of the neutral gas. In particular,
if no large obscuration is found against the very central core also from the
HI absorption, the lack or weakness of broad optical lines in FRI radio
galaxies compared to other AGN will have to be explained by something other
than obscuration effects.  So far, broad lines have been {\sl tentatively }
found only in very few cases of FRI radio galaxies.

\section{Interpreting the HI absorption: the problem of the redshift}

An important issue for understanding the origin of the \HI\ absorption is how
the systemic velocity of the galaxy compares with the velocity of the \HI.  It
was already pointed out by Mirabel (1989) how the systemic velocities derived
from emission lines can be both uncertain and biased by motions of the
emitting gas. This is nicely illustrated by our results on the compact radio
galaxy PKS 1549-79 (see Tadhunter et al. 2001 for details).  At radio
wavelengths, PKS~1549--79 is a compact (the size is 150 mas, about 350 pc for $H_\circ = 50$ \kms Mpc$^{-1}$ and
$q_\circ = 0$)  flat spectrum source with a one-sided jet.  The radio
structure indicates that the radio jet axis is aligned with our line of sight.
Despite this, HI absorption was detected (optical depth $\sim 2$ \%) and
the profile is shown in Fig. 2.  In the optical, the most surprising
characteristics is that {\sl two
  redshift systems} were found from the emission lines: the higher ionisation
lines (e.g.\ [OIII]5007\AA) have a significant lower redshift (velocity
difference of $\Delta v = 600$ \kms, see Fig.\ 2) than the low ionisation
lines (e.g.\ [OII]3727\AA). {\sl The velocity derived from the low ionisation
  lines is consistent with the one derived from the \slHI}. {\sl No evidence
  for broad permitted lines} has been found, quite unusual for a flat spectrum
radio source and the  high ionisation forbidden lines are also unusually
broad (1345 \kms).

\begin{figure}[!h]
\centerline{\psfig{figure=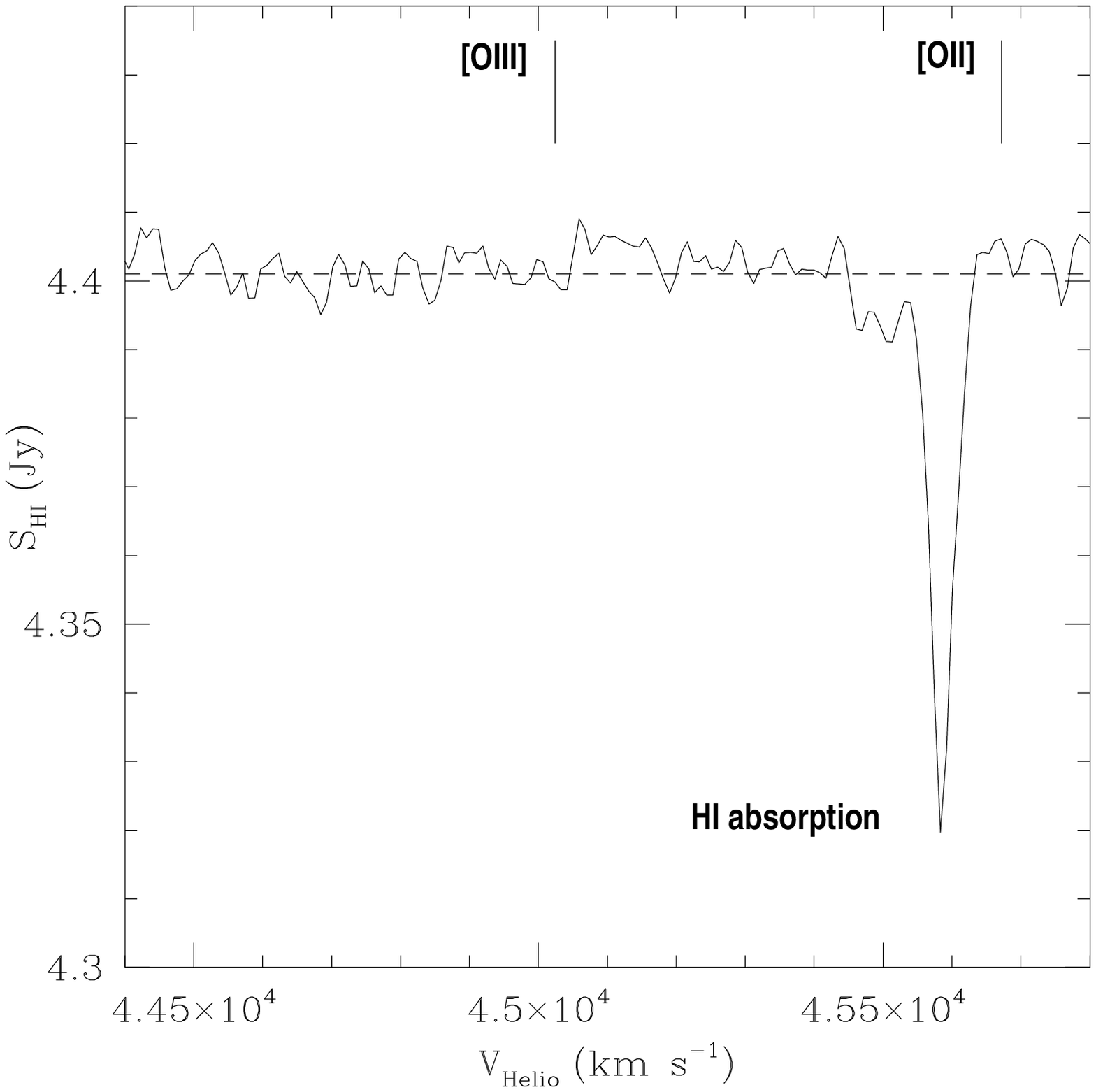,width=5cm}
\psfig{figure=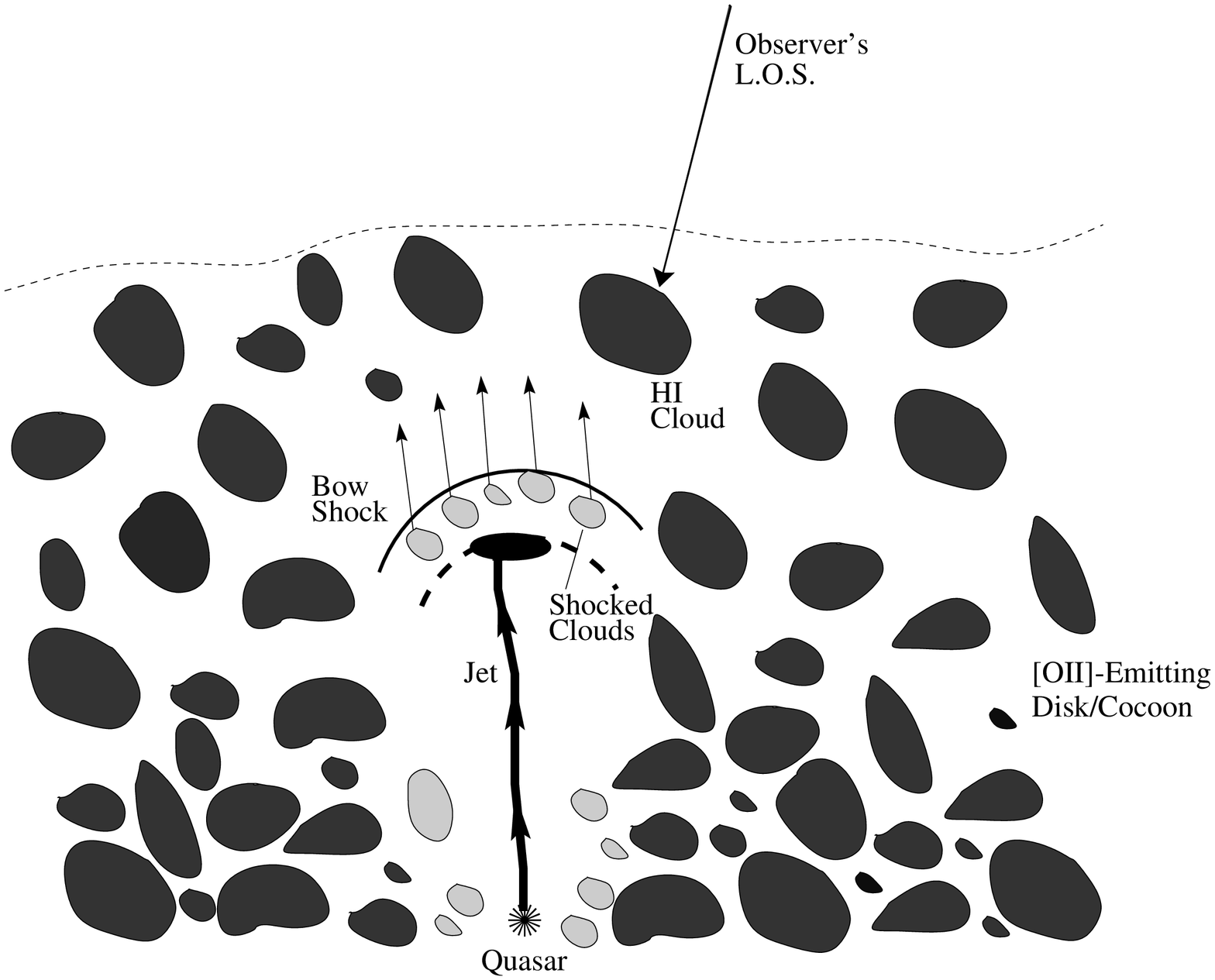,width=7.5cm,angle=-90}}
\caption{{\sl (Left)} HI absorption profile obtained with the Australian LBA of PKS~1549-79 with  superimposed the velocity of the ionised gas. {\sl (Right)}
  Cartoon of the possible distribution of the various components (see text for
  details).}
\end{figure}

All this suggests that in PKS~1549--79 the high ionisation lines are formed in
a region close to the central AGN, which is undergoing outflow because of
interactions, e.g., with the radio jet, while the low ionisation lines and the
HI absorption come from an obscuring region at larger distance and not so
disturbed kinematically.  Moreover, PKS~1549--79 is one of the few known radio
galaxies at low redshift which shows a strong component from a young stellar
population (see \S 5 for some more on these galaxies), and strong far-IR
emission. Thus, it is likely that it is a {\sl young} source where the nucleus
is surrounded by a cocoon of material left over from the events which
triggered the nuclear activity.  As the radio source evolves, any obscuring
material along the jet axis is likely to be swept aside by, e.g., jet-cloud
interactions.  In Fig. 2 is shown a schematic diagram of the various
components.
A major implication of this work is that the simplest version of the unified
schemes {\sl may not always hold for young, compact radio sources}.
The possibility that  the medium around (i.e. in the
form of a starburst component) could affect the way we see the AGN especially
at optical wavelengths has been recently pointed out also by the work of
Levenson et al. (these proceedings) in the case of the
Starburst/Seyfert-1 galaxy NGC~6221.


\section{Infall and outflow}

In previous studies, the HI absorption was mainly found either at the systemic
velocity or redshifted compared to it (van Gorkom et al. 1990).
As a result of the HI studies we have done so far, we found both redshifted and
blueshifted (with respect to the systemic velocity) cases of HI absorption.
However, it is clear from the above discussion that, in general, accurate
redshifts are needed before drawing any strong conclusion.

At present, there are only a few cases of clearly redshifted HI absorption where
a cloud falling into the nucleus could be the cause for the absorption. One of
these cases is NGC~315. As evident from Fig. 1, two HI absorption systems are
detected: a newly discovered (broad) HI absorption at the systemic velocity
and a well known deep and very narrow component (Dressel et al. 1983) 500 km/s
redshifted compared to the systemic velocity. Similar double HI absorptions
have been found only in NGC~1275 and 4C~31.04.  The real cause of the
redshifted absorption  is
still unclear. Despite our very sensitive HI observation done with the WSRT,
we detect this absorption {\sl only} against the nucleus (no similar
absorption at the 5\% optical depth limit is seen against, e.g., the radio
jet).  This may mean that is due to a cloud physically associated with the
nuclear regions, i.e. a cloud close to the nucleus, falling into it and "feeding" the AGN.


In some cases, the HI absorption seems to come from gas situated around the
radio lobes affected by the interaction with the radio plasma (i.e. outflow).
The best example is the Seyfert galaxy IC~5063.  In this galaxy, very broad,
blue-shifted (up to $\sim 700$ \kms) HI absorption has been detected
(Oosterloo et al. 2000) at the location of the western radio hot spot at about
700 pc from the nucleus. Such velocities are much too high to be explained by
gravitational motion and the only viable explanation is that at the location
of the hot spot the radio jet is hitting the ISM of the galaxy, pushing out
the ISM at these high velocities.
An other example is the compact radio galaxy PKS 1814-63. In this galaxy the
HI absorption is observed against the entire radio emission (Morganti et al.
2000). Most of the HI absorption (with optical depth as deep as 30\%) is
blueshifted compared to the systemic velocity of the galaxy (at least if we
rely on the redshift available so far).  Thus, this component could be associated
with extended gas, possibly surrounding the lobes and perhaps
interacting/expanding with them.

\section{Starburst radio galaxies}

A small fraction of powerful radio galaxies shows strong optical/UV starburst
activity as well as strong far-IR emission (Wills et al. 2001 and these proceedings).  For these
objects it is generally assumed that the far-IR excess represents reprocessed
starlight.  These powerful radio galaxies are potentially valuable probes of
the formation and evolution of radio galaxies and QSRs, as well as early-type
galaxies in general.

Although the statistics is limited, we found a tendency for the far-IR bright,
starbursting radio galaxies to be detected in HI absorption.  Apart from the
two objects mentioned above (PKS 1549-79 and 4C12.50) we have detected HI
absorption in 3C459 and B2 0648+27 while other cases were already known from
the literature (3C321 and 3C433 Mirabel 1989, 1990).  Thus, the rich ISM
likely to be present in these galaxies, as suggested by the strong far-IR
emission and the star formation, could have some relation with the detection
of HI absorption.  In some of these cases, HI absorption can be due to a
nuclear torus/disk.  An important part of the process of formation of radio
galaxies described above would be the formation of a self-graviting gas disk
that would bring the gas into the nucleus and fuel the AGN. Evidence for a
disk has been found in some ultraluminous infrared galaxies (ULIG) and a
connection between ULIGs and powerful radio galaxies with starburst activity
and far-IR emission could exist in that the latter could represent the first
evolutionary result of ULIG.  However, in other cases, gas in turbulent motion
or tails of gas remnant from the formation of the galaxy itself could have a
role in producing the absorption.  This could be the case in PKS 1549-79, as
described above, but also in other objects (like 3C433) where the absorption
is seen against one of the radio lobes instead of the nucleus.

\section{Summary}

We have presented some results from studies of HI absorption in radio galaxies.
We have found that the absorption can be due to a variety of phenomena.  In
observed the FRI radio galaxies, the detected absorption is likely to be due to circumnuclear
disk/torus that could be geometrically thin, consistent with what
derived from HST data.  However, many other situations have been found:
extreme outflows, more quiescent absorbing clouds, and structures likely to be
associated with a merger remnant.  In particular, we find a tendency for radio
galaxies with a strong component of young stellar population and far-IR
emission to show HI absorption.  Thus, HI absorption gives a powerful tool to
probe different phenomena in and around radio loud AGNs.



\begin{references}


\reference Capetti, A., de Ruiter, H. R., Fanti, R., Morganti, R., Parma, P.,
\& Ulrich, M.-H.  2000, \aap, 362, 871
\reference Chiaberge et al. 1999, A\&A 349, 77;
\reference Conway J.E.  1998 in {\sl Highly redshifted radio lines},
Carilli C.L. et al.  eds., ASP Conf.  Series 156, 259;
\reference Dressel L., Bania T.M., Davis M.M. 1983, ApJ 266, L97
\reference Gunn J.E. 1979, in {\sl Active Galactic Nuclei}, Hazard et al. eds,
CUP p.213
\reference Maloney P.R., Hollenbach D.J., Tielens A.G.G.M. 1996, ApJ 466, 561
\reference Mirabel I.F. 1990, ApJ 352, L37
\reference Mirabel I.F. 1989, ApJ 340, L13
\reference Morganti R. et al. 2000,
'5th European VLBI Network Symposium' 
Conway J.E. et al. eds,  ISBN 91-631-0548-9, p. 111 (astro-ph/0010482)
\reference Morganti et al. 2001, MNRAS 323, 331
\reference Oosterloo et al. 2000, AJ 119, 2085
\reference van Langevelde H.J. et al. 2000, A\&A 354, 45
\reference Tadhunter C., Wills K., Morganti R., Oosterloo T., Dickson R.,
2001, MNRAS in press (astro-ph/0105146)
\reference Taylor G.B. 1996, ApJ 470, 394
\reference van Gorkom J.H., Knapp G.R., Ekers R.D., Ekers D.D., Laing R.A.  \&
Polk K., 1989, AJ 97, 708
\reference Wills K.A., Tadhunter C.N., Robinson T.G., Morganti R. 2001, MNRAS submitted

\end{references}
\end{document}